# Electronic states and magnetic properties of edge-sharing Cu-O chains

Y. Mizuno*, T. Tohyama, and S. Maekawa

*Institute for Materials Research, Tohoku University, Sendai 980-77, Japan*

T. Osafune, N. Motoyama, H. Eisaki, and S. Uchida

*Department of Superconductivity, The University of Tokyo, Bunkyo-ku, Tokyo 113, Japan*

(January 20, 1998)

The electronic states and magnetic properties for the copper oxides containing edge-sharing Cu-O chains such as $Li_2CuO_2$, $La_6Ca_8Cu_{24}O_{41}$ and $CuGeO_3$ are systematically studied. The optical conductivity $\sigma(\omega)$ and the temperature dependence of the magnetic susceptibility $\chi(T)$ for single crystalline samples $Li_2CuO_2$ are measured as a reference system and analyzed by using the exact diagonalization method for small Cu-O clusters. It is shown that the spectral distribution of $\sigma(\omega)$ is different between edge-sharing and corner-sharing Cu-O-Cu bonds. The charge transfer gap in edge-sharing chains is larger than that of high-$T_c$ cuprates. The exchange interaction between nearest-neighbor copper ions in edge-sharing chains $J_1$ depends sensitively on the Cu-O-Cu bond angles. In addition to $J_1$, the exchange interaction between next-nearest-neighbor copper ions $J_2$ has sufficient contribution to the magnetic properties. We calculate $J_1$ and $J_2$ for all the copper oxides contaning edge-sharing Cu-O chains and discuss the magnetic properties.

PACS appear here. \pacs{} should always be input, even if empty.

## I. INTRODUCTION

The physical properties of the one-dimensional (1D) copper oxides have drawn much attention in connection with the physics of high-$T_c$ cuprates. The 1D copper oxides are classified by the structure into two types. One is corner-sharing Cu-O chain where $CuO_4$ units share their corners to form the chain. Typical examples are $Sr_2CuO_3$ and $SrCuO_2$. High-$T_c$ cuprates are of two-dimensional (2D) version of this type. The other is edge-sharing chain where $CuO_4$ units share their edges. Since the edge-sharing Cu-O chain has nearly 90° Cu-O-Cu bonds, different behaviors of electronic states and magnetic properties are expected compared to the corner-sharing Cu-O chain with 180° Cu-O-Cu bonds.

The most important feature of the edge-sharing Cu-O chain is that an $O2p_\sigma$ orbital hybridizing with a $3d$ orbital of Cu ion is almost orthogonal to that of the next Cu ion. In contrast with this case, in the corner-sharing chain a $2p_\sigma$ orbital hybridizes with the two neighboring $Cu3d$ orbitals. This implies that electronic states of the two types of chain behave differently as far as the propagation of carriers is concerned. The difference can be directly reflected by, for example, the optical conductivity $\sigma(\omega)$. The orthogonality of the orbitals also affects magnetic interaction in the chain. According to the Goodenough-Kanamori-Anderson rule,[1-3] the nearest-neighbor (NN) Cu-Cu spin interaction, $J_1$, changes from antiferromagnetic (AFM) to ferromagnetic one (FM), as the angle $\theta$ of Cu-O-Cu bond approaches 90 degrees. This is because the AFM superexchange interaction is reduced by the orthogonality. In addition to $J_1$, the next-nearest-neighbor (NNN) interaction, $J_2$, with AFM contribution through Cu-O-O-Cu paths, is expected to exist with considerable magnitude.

Although such intuitive arguement is known for the edge-sharing Cu-O bond, there are few systematic investigations of nature of the bond. The only study has been performed on di-$\mu$-hydroxo-bridged copper (II) complexes.[4] However, the complexes have not chains, but isolated Cu-O-Cu bridges. In this paper, we present a systematic investigation of electronic and magnetic properties for the edge-sharing Cu-O chain based on recent progress of the study of copper oxides, together with new experimental results.

As typical insulating materials which contain the edge-sharing chains, $La_6Ca_8Cu_{24}O_{41}$[5,6] and $CuGeO_3$[7] are known, and have intensively been studied. The former material consists of three layers: (i) the $Cu_2O_3$-ladder layer, (ii) the (La,Ca) layer, and (iii) the edge-sharing $CuO_2$-chain layer. The angle $\theta$ of the chain is 91 degrees. This is also known as parent compound of $Sr_{14-x}Ca_xCu_{24}O_{41}$ that becomes superconducting at $x=13.6$[8] and $11.5$[9] under high pressure. The crystal structure is shown in Fig.1(a). The latter material shows the spin-Peierls transition at around 14 K.[7] The crystal structure is depicted in Fig.1(b), and $\theta$ is about 99 degrees. Both materials are not simple chain systems: $La_6Ca_8Cu_{24}O_{41}$ has not only chains, but also ladders. In $CuGeO_3$, chains are tilting alternately about 60 degrees, and as a result $Ge^{4+}$ ions are very close to in-chain oxygen.

Another material with edge-sharing chains is $Li_2CuO_2$. The crystal structure is shown in Fig.1(c). The angle $\theta$ is 94 degrees. This material consists of a simple edge-sharing chains situated parallel to each other. $Li_2CuO_2$ is, therefore, thought to be an ideal candidate for the edge-sharing Cu-O chain. The examination of the electronic states of $Li_2CuO_2$ is very important in order to clarify the characteristics of the electronic states of the edge-sharing chain.

In previous works on $Li_2CuO_2$, the temperature dependence of the magnetic susceptibility $\chi(T)$ and powder neutron diffraction measurements have been performed for polycrystalline samples.[10,11] Three-dimensional (3D) AFM ordering occurs at $T_N = 8.3$ K.[10] The susceptibility data above $T_N$ follow the Curie-Weiss law with negative Weiss temperature. This indicates AFM interaction between NN Cu spins. On the other hand, the neutron data have been analyzed assuming FM NN interaction together with FM and AFM interchain interactions along a- and c-axes, respectively. In both analyses, $J_2$ was ignored. Since any consistent analyses among experiments have not been done, further examination for $Li_2CuO_2$ is necessary both experimentally and theoretically.

In this study, we first report new experimental data of the optical conductivity $\sigma(\omega)$ and $\chi(T)$ for single crystalline sample of $Li_2CuO_2$. Next, in order to analyze the experimental results, the electronic states for $Li_2CuO_2$ are examined based on the ionic and cluster model approaches which were successfully applied to the study of high $T_c$ cuprates and ladder compounds.[12–14] Lastly, by using the same procedure used in analyses of $Li_2CuO_2$, the electronic states and the magnetic properties of $La_6Ca_8Cu_{24}O_{41}$ and $CuGeO_3$ are investigated. In addition to these materials, ladder materials which contain the edge-sharing structure are examined. In particular, we calculate the interladder magnetic interactions for the ladder materials. Thereby, the systematics of electronic and magnetic properties for the edge-sharing Cu-O chains is clarified.

A brief summary of this paper is shown below.

Since the present systems are the charge transfer (CT)-type insulators, the CT energy $\Delta$, which is the energy-level separation between Cu$3d$ and O$2p$ orbitals, plays an important role. $\Delta$ is dependent on (i) the difference of Madelung potential between Cu and O sites and (ii) the polarization effect represented by the dielectric constant. The calculated values of $\Delta$ for $Li_2CuO_2$ and $La_8Ca_6Cu_{24}O_{41}$ are comparable with those for high $T_c$ cuprates. However, since small polarization effect is expected in the chain system, the resulting insulating gap (CT gap) is larger (2.1~2.2 eV) compared to high $T_c$ cuprates (1.5~2 eV). The difference of Madelung potential for $CuGeO_3$ is the largest among those for all the cuprates studied so far. This is because of the presence of Ge ions close to the O ions. The estimated CT gap is about 3.2 eV. This is consistent with recent optical measurements.[15,16]

The spectral distribution of $\sigma(\omega)$ is in good accord with calculated $\sigma(\omega)$ for a small cluster by using the exact diagonalization method. The weight of a peak called CT peak is strongly suppressed due to the orthogonality of the orbitals, because the weight is related to the propagation of an excited hole. On the other hand, higher energy excitations from nonbonding O$2p$ band to Cu$3d$ upper Hubbard band give strong intensity in $\sigma(\omega)$. These behaviors are in contrast with those for the 180° Cu-O-Cu bond.

The magnetic interaction between NN Cu spins, $J_1$, shows a systematic feature as a function of the angle $\theta$ as expected. $J_1$ is FM for $La_8Ca_6Cu_{24}O_{41}$ ($\theta = 91°$) and $Li_2CuO_2$ ($\theta = 94°$) but AFM for $CuGeO_3$ ($\theta = 99°$). The NNN interaction $J_2$ is always AFM and is not much dependent on $\theta$. From the analyses of $\chi(T)$, it is found that $J_2$ is important in order to understand the magnetic properties of the edge-sharing Cu-O chain. By using the theoretically estimated values of $J_1$ and $J_2$, we reproduce the experimental data of $\chi(T)$ for $Li_2CuO_2$ and $La_8Ca_6Cu_{24}O_{41}$. However, in $CuGeO_3$ the agreement is not excellent between theory and experiment. Our cluster calculation gives the ratio $\alpha(\equiv J_2/J_1)=0.13$ for $CuGeO_3$, which is smaller than that obtained from the previous studies.[17–20]

The outline of the paper is as follows. The general features of the electronic and magnetic properties in the edge-sharing Cu-O chain are summarized in Sec. II to make underlying physics clear. In Sec. III, the experimental result of $\sigma(\omega)$ and $\chi(T)$ for single crystalline $Li_2CuO_2$ is presented. In Sec. IV, theoretical analyses for $\sigma(\omega)$ are performed using a small cluster. Good agreement between experiment and theory is obtained. In addition, the magnetic interactions, $J_1$ and $J_2$, are evaluated with help of experimental data of $\chi(T)$ for $Li_2CuO_2$. By using the obtained values of $J_1$ and $J_2$, $\chi(T)$ for $La_8Ca_6Cu_{24}O_{41}$ is reproduced, but not for $CuGeO_3$. The last section (Sec. V) is devoted to discussions showing interladder interactions for ladder materials.

## II. GENERAL FEATURES IN THE EDGE-SHARING CHAINS

In this section, we discuss general features of the electronic and magnetic properties of the edge-sharing Cu-O chains. The edge-sharing chains are formed by the $CuO_4$ units which share their edges. The angle for Cu-O-Cu bonding, $\theta$, is nearly 90 degrees (Fig. 2(a)). When the axes of coordinates are taken along Cu-O bonds ($x'$-$y'$ axes) as shown in Fig. 2(b), the $d_{x'^2-y'^2}$ orbital on Cu ion strongly overlaps with the two $2p$ orbitals ($p_{x'}$ and $p_{y'}$) on neighboring O ions, but does not with the other orbital. Therefore, it is apparent that the effective hopping between neighboring Cu$d_{x'^2-y'^2}$ orbitals through O ions is rather small if $\theta$ is close to 90 degrees. The vanishing of the effective hopping is characteristic of the edge-sharing chains and is in strong contrast to the corner-sharing chain where the neigboring Cu$d_{x^2-y^2}$ orbitals are connected via the same O$2p_\sigma$ orbital.

In the following, we use new axes of coordinates rotated by 45 degrees from the $x'$-$y'$ axes (Fig. 2(c)). The model Hamiltonian suitable for describing the electronic structure of the edge-sharing Cu-O chain is the three band Hubbard Hamiltonian,

$$H = \sum_{m,l,\alpha,\sigma} T_{dp_\alpha}^{ml}(d_{m\sigma}^\dagger p_{\alpha l\sigma} + h.c.) + \sum_{l,l',\alpha,\sigma} T_{p_\alpha p_\alpha}^{ll'} p_{\alpha l\sigma}^\dagger p_{\alpha l'\sigma}$$

$$+ U_d \sum_m n_{m\uparrow}^d n_{m\downarrow}^d + U_p \sum_{l,\alpha} n_{l\uparrow}^{p_\alpha} n_{l\downarrow}^{p_\alpha}$$

$$+ \sum_{\sigma\sigma'} (U_{pp} - \delta_{\sigma\sigma'} K_p) \sum_l n_{l\sigma}^{p_x} n_{l\sigma'}^{p_y}$$

$$- K_p \sum_{l,\sigma} p_{xl\sigma}^\dagger p_{xl\bar\sigma} p_{yl\bar\sigma}^\dagger p_{yl\sigma}$$

$$+ K_p \sum_l (p_{xl\uparrow}^\dagger p_{yl\uparrow} p_{xl\downarrow}^\dagger p_{yl\downarrow} + p_{yl\uparrow}^\dagger p_{xl\uparrow} p_{yl\downarrow}^\dagger p_{xl\downarrow}) \quad (1)$$

$$+ \sum_{\sigma\sigma'} (U_{pd} - \delta_{\sigma\sigma'} K_{pd}) \sum_{m,l,\alpha} n_{m\sigma}^d n_{l\sigma'}^{p_\alpha}$$

$$- K_{pd} \sum_{m,l,\alpha,\sigma} d_{m\sigma}^\dagger d_{m\bar\sigma} p_{\alpha l\bar\sigma}^\dagger p_{\alpha l\sigma}$$

$$+ K_{pd} \sum_{l,m,\alpha} (d_{m\uparrow}^\dagger p_{\alpha l\uparrow} d_{m\downarrow}^\dagger p_{\alpha l\downarrow} + p_{\alpha l\uparrow}^\dagger d_{m\uparrow} p_{\alpha l\downarrow}^\dagger d_{m\downarrow}) \quad (2)$$

$$+ \epsilon_d \sum_{m,\sigma} n_{m\sigma}^d + \epsilon_p \sum_{l,\alpha,\sigma} n_{l\sigma}^{p_\alpha}, \quad (3)$$

where the operators $d_{m\sigma}^\dagger$ and $p_{\alpha l\sigma}^\dagger$ create holes in Cu$3d_{xy}$ and O$2p_\alpha$ ($\alpha = x$ or $y$) orbitals with spin $\sigma(=\uparrow$ or $\downarrow)$ at the $m$ and $l$ sites, respectively. $n_{m\sigma}^d = d_{m\sigma}^\dagger d_{m\sigma}$ and $n_{l\sigma}^{p_\alpha} = p_{\alpha l\sigma}^\dagger p_{\alpha l\sigma}$. The $m$, $l$ and $l'$ summations are over the orbitals at the neighbor sites. $T_{dp_\alpha}^{ml}$ and $T_{p_\alpha p_\alpha}^{ll'}$ are the hopping parameters between the neighboring $3d_{xy}$ and $2p_\alpha$ orbitals and between the neighboring $2p_\alpha$ orbitals, respectively. They are determined according to the relation given by Slater and Koster,[21] i.e., $T_{dp_x}^{ml} = \sqrt{3}\zeta^2\eta(pd\sigma) + \eta(1-2\zeta^2)(pd\pi)$, $T_{dp_y}^{ml} = \sqrt{3}\eta^2\zeta(pd\sigma) + \zeta(1-2\eta^2)(pd\pi)$, $T_{p_\alpha p_\alpha}^{ll'} = -(pp\sigma)$ for $\sigma$ bond and $T_{p_\alpha p_\alpha}^{ll'} = (pp\pi)(=-0.25(pp\sigma)^{22})$ for $\pi$ bond, where $(pd\sigma)$, $(pd\pi)$ and $(pp\sigma)$ are the hopping integral parameters which are dependent on bond length, and $\zeta$ and $\eta$ denote the $x$ and $y$ components of the cosines of the direction for a vector connecting $m$ and $l$ sites. Note that $T_{dp_x}^{ml} = T_{dp_y}^{ml}$ for $\theta = 90°$ because $|\zeta| = |\eta| = 1/\sqrt{2}$, while $T_{dp_x}^{ml} > T_{dp_y}^{ml}$ for $\theta > 90°$. $U_d$ and $U_p$ represent the intra-orbital Coulomb interactions on $3d_{xy}$ and $2p_\alpha$ orbitals, respectively. $K_p$ represents Hund's coupling on the $2p_\alpha$ orbitals. $K_{pd}$ is the direct exchange interaction between neighboring $3d_{xy}$ and $2p_\alpha$. $U_{pp}$ and $U_{pd}$ denote the inter-orbital Coulomb interactions between $2p_x$ and $2p_y$ and between $3d_{xy}$ and $2p_\alpha$, respectively. $U_{pp}$ is given by $U_p - 2K_p$. The energy difference of the on-site potentials, $\epsilon_p - \epsilon_d$, is called as the CT energy $\Delta$.

Insulating copper oxides are classified as the CT type ($\Delta < U_d$). In Fig. 3, the schematic density of states for the CT insulator is shown. In the photoemission process, a photo-doped hole entered into O$2p$ orbitals is coupled to a Cu$3d$ hole due to the strong hybridization. As a result, a bound state called the Zhang-Rice (ZR) local singlet[23] is formed, and is split from oxygen band (non-bonding (NB) band). The gap energy is determined by the inter-band excitation from the ZR singlet band to the upper Hubbard (UH) band denoted by process **A**. In $\sigma(\omega)$, this process appears as the lowest-energy excitation. This is called the CT excitation. In addition, another excitation is seen in the higher energy region of $\sigma(\omega)$, which corresponds to the excitation from the O$2p$ NB bands to the UH bands represented by process **B**. Note that, if we use a single CuO$_4$ cluster in evaluating $\sigma(\omega)$, only this excitation can be seen in $\sigma(\omega)$. In this sense, this is a local excitation. These two excitations, **A** and **B**, are characteristic of the CT insulator and are seen in both corner- and edge-sharing chains. An important difference between the two types of chain appears on the distribution of the weights of these excitations in $\sigma(\omega)$. The intensity of the CT excitation for the edge-sharing chains is small compared to that for the corner-sharing chains. This is because the hole propagation to neighboring CuO$_4$ units is suppressed due to the near 90° Cu-O bonding. If $\theta = 90°$ and $T_{p_\alpha p_\alpha}^{ll'} = 0$, the intensity of the CT excitation vanishes completely. On the other hand, the intensity due to the process **B** is stronger in the edge-sharing chain than in the corner-sharing chain because the $\omega$-sum of the weights is nearly conserved between the two types of chain. Thus, we can directly observe the difference of the electronic states by investigating $\sigma(\omega)$ (see Sec. III).

In the insulating cuprates, each Cu site possesses a localized spin with $S = 1/2$. The nature of exchange interaction between the Cu spins is different from that of the corner-sharing chains due to the near 90° Cu-O bonding. In the corner-sharing chains, the NN interaction $J_1$ is AFM with the magnitude of more than 2000 K.[24,25] On the contrary, $J_1$ for the edge-sharing chain is AFM or FM depending on the angle $\theta$. Furthermore, its magnitude is expected to be fairly small. Here, we briefly discuss the microscopic origin of $J_1$. $J_1$ is expressed as the sum of FM and AFM contributions. The FM contribution is caused by (i) the direct exchange interaction between Cu$3d$ and O$2p$ orbitals, $K_{pd}$, and (ii) Hund's coupling on two O$2p$ orbitals, $K_p$. On the other hand, the AFM contribution is due to Anderson's superexchange process where the hybridization between Cu$3d$ and O$2p$ plays an important role. If $\theta = 90°$ (and $T_{p_\alpha p_\alpha}^{ll'} = 0$), this AFM process is impossible because of the absence of common O$2p$ orbital combining a Cu ion with neigboring Cu ions. In this case $J_1$ becomes FM. With increasing $\theta$ from 90 degrees, the AFM contribution increases, but net interaction is still FM. At a special angle of $\theta$, the FM contribution is completely compensated for by the AFM one and substantially $J_1$ becomes zero. With further increase of $\theta$, $J_1$ becomes AFM. Such behaviors are known as the Goodenough-Kanamori-Anderson rule.[1–3]

It is clear from Fig. 2(c) that NNN CuO$_4$ units are connected through the hybridization of NN O$2p_x$ orbitals. Since the corresponding hopping parameter $T_{p_x p_x}^{ll'}$ is not small, there is an exchange interaction, $J_2$, between the NNN Cu spins through Cu-O-O-Cu path. $J_2$ is AFM and gives the effect of the frustration into the chain. The magnitude of $J_2$ is expected to be comparable with that of $J_1$ under some conditions. Thus, the spin Hamiltonian for the edge-sharing chains can be written as

$$H = \sum_i (J_1 \mathbf{S}_i \cdot \mathbf{S}_{i+1} + J_2 \mathbf{S}_i \cdot \mathbf{S}_{i+2}). \quad (4)$$

It has been known that this model contains various phases depending on the magnitude of $J_1$ and $J_2$.[26-30] This means that various magnetic properties can be seen in the materials with the edge-sharing chains.

## III. EXPERIMENTAL RESULTS OF $LI_2CUO_2$: OPTICAL CONDUCTIVITY & MAGNETIC SUSCEPTIBILITY

In this section, we show experimantal data of the optical conductivity $\sigma(\omega)$ and the magnetic susceptibility $\chi(T)$ for $Li_2CuO_2$. Single-crystalline samples of $Li_2CuO_2$ were grown by the traveling-solvent floating zone (TSFZ) method. The $CuO_4$ units are lying on the b-c plane and are linked together along the b-axis. The resulting $CuO_2$ chains are stacked in the a-axis direction (see Fig. 1(c)). Reflectivity was measured at room temperature using polarized light in the energy region 0.01–40 eV at near-normal incidence to the b-c plane. $\sigma(\omega)$ was obtained by the Kramers-Kronig transformation made on the reflectivity spectrum. In Fig. 4, the experimental data of $\sigma(\omega)$ with the polarization parallel to the chain (b-axis) for $Li_2CuO_2$ is shown. The spectrum has a peak at around $\omega=4$ eV and the intensity of the spectrum decreases monotonously toward the low-energy region. This behavior is in strikingly contrast to those of the corner-sharing chains[31] and planes[32] showing CT peaks at around 2 eV related to the CT excitation mentioned in Sec. II.

The magnetic susceptibility measurements for single crystalline $Li_2CuO_2$ were done using a superconducting quantum interface device (SQUID) dc magnetometer between 2 and 400 K. In Fig. 5, the experimental result in the magnetic field applied along the three principal axes is shown. In the neutron diffraction experiment the AFM ordering has been pointed out.[10] The susceptibility with the magnetic field along a-axis shows the adrupt depression below 9 K. The susceptibilities with the magnetic field along b- and c-axis are almost constant with increasing temperature up to 9 K, above which they decrease with increasing temperature. These indicate that the AFM transition occurs below 9 K. This behavior is consistent with the spin structure proposed by neutron powder diffraction[10] and paramagnetic and antiparamagnetic resonance experiments[33].

## IV. THEORETICAL RESULTS

### A. Optical Conductivity

We perform theoretical investigation on $\sigma(\omega)$ for $Li_2CuO_4$ by the ionic and cluster model approaches. The model Hamiltonian for the edge-sharing chains was shown in Eq. (1). Within the ionic model, the CT energy $\Delta$ can be expressed as[12]

$$\Delta = \Delta V_M / \varepsilon_\infty + \Delta_0, \quad (5)$$

where $\Delta V_M$ is the difference in the Madelung site potential for a hole between Cu and O sites, $\varepsilon_\infty$ is the dielectric constant, and $\Delta_0$ is related to the second ionization energy of a $Cu^{2+}$ ion and the second electron affinity of an $O^{2-}$ ion. $\Delta V_M$ is calculated using the crystal structure for $Li_2CuO_2$.[10] $\Delta_0$ is taken to be $-10.88$ eV as for other high-Tc cuprates.[12] $\varepsilon_\infty$ is set to be 3.3, which is slightly smaller than 3.5 for $CuO_2$ plane.[12] This is because the number of neighboring oxygens around a oxygen is smaller than that for the $CuO_2$ plane. As standard parameters of the hopping integrals, we choose $(pd\sigma)=1.30$ eV at $d_{Cu-O}=1.905$ Å and $(pp\sigma)=-0.59$ eV at $d_{O-O}=2.808$ Å. The former value was used in previous studies for $La_2CuO_4$.[12] The latter one was taken from angle-resolved photoemission data for an insulating cuprate $Sr_2CuO_2Cl_2$.[34] To obtain hopping integral parameters for $Li_2CuO_2$, we assume the bond length dependences with $d^{-4}$ and $d^{-3}$ for $(pd\sigma)$ and $(pp\sigma)$, respectively. The resulting values are listed in Table I. The on-site Coulomb energies are set to be $U_d=8.5$ eV and $U_p=4.1$ eV as for the previous studies.[12] The Hund's coupling on O ion is assumed to be $K_p=0.6$ eV, and the direct exchange interaction between $Cu3d_{xy}$ and $O2p_\alpha$ is taken to be $K_{pd}=0.05$ eV. The latter value is determined by fitting theoretical $\chi(T)$ to experimental one measured for the single crystalline sample, as will be discussed later. Note that these Coulomb interactions are insensitive to the structure of $\sigma(\omega)$ at the energy region where we are interested. The inter-site Coulomb interation $U_{pd}$ is neglected in this study because part of the effect of $U_{pd}$ is included in the level-separation $\Delta$. The neglect of $U_{pd}$ coresponds to ignoring excitonic effect in $\sigma(\omega)$. However, since the excionic effect does not produce qualitative changes on the systematic evolution of the electronic states which we are interested in, our conclusions are not affected by setting $U_{pd}=0$.

The exact diagonalization method is used for the calculation of $\sigma(\omega)$ for the $Cu_4O_{10}$ cluster with free boundary condition. The real part of $\sigma(\omega)$ is given by the Kubo formula ($\hbar = c = 1$),

$$\sigma(\omega) = \frac{\pi}{\omega} \sum_n |\langle n | J_x | 0 \rangle|^2 \delta(\omega - E_n + E_0), \quad (6)$$

where $|0\rangle$ and $|n\rangle$ are the ground and excitaed states with energies $E_0$ and $E_n$, respectively. The light is applied to the direction parallel to the chain. The operator $J_x$ is the x-component of the current operator $\hat{J}$,

$$\hat{J} = ie\left[\hat{d}_{\text{Cu-O}}\sum_{m,l,\sigma}\left(p^{\dagger}_{\alpha l\sigma}d_{m\sigma} - d^{\dagger}_{m\sigma}p_{\alpha l\sigma}\right) + \hat{d}_{\text{O-O}}\sum_{l,l',\alpha,\sigma}p^{\dagger}_{\alpha l\sigma}p_{\alpha l'\sigma}\right], \quad (7)$$

where $e$ is the electric charge and the vectors $\hat{d}_{\text{Cu-O}}$ ($\hat{d}_{\text{O-O}}$) connect $m$ and $l$ ($l'$ and $l$) sites. Equation (6) is evaluated by using the Lanczos algorithm.

Calculated spectrum for $\text{Li}_2\text{CuO}_2$ is shown in Fig. 6(a) with the Lorentzian broadening with 0.6 eV half width. The result is in good agreement with the experimental one shown in Fig. 4. The peak existing at around 4 eV has strong intensity, which corresponds to the excitation from the O2$p$ NB bands to the UH bands as denoted by **B** in Fig. 3. Even for other cuprates with corner-sharing chains and planes, such an excitation has been observed theoretically[35] and experimentally[36] at the same energy region ($4 \leq \omega \leq 5$ eV). This excitation is, therefore, independent of the type of Cu-O bond. In other words, this is related to the local excitation within a CuO$_4$ unit. The magnitude of CT gap is determined from the excitation process of **A** denoted in Fig. 3. In the calculated result, structure due to process **A** appears at around 2.2 eV with very weak intensity. This is characteristic of the edge-sharing Cu-O chain as mentioned in Sec. II. In the experimental result, this structure can not be seen probably due to weak intensity. We therefore conclude that the experimental result of $\sigma(\omega)$ for $\text{Li}_2\text{CuO}_2$ captures the distinctive features of the electronic states ascribed to the structure of the edge-sharing chain. Here, we note that the calculated CT gap of 2.2 eV is somewhat larger than those for high-$T_c$ cuprates (1.5~2 eV). In general, experimentally obtained CT gap is larger for 1D cuprates than those for 2D cuprates.[31] Since the contribution of the Madelung potentials $\Delta V_M$ to $\Delta$ is similar between 1D and 2D cuprates, the reduction of dielectric effect due to small number of oxygen ions is essential to produce the large CT gap (see Eq (5)). Finally we emphasize that the good agreement of the theoretical $\sigma(\omega)$ with the experimental one assures us of the validity of our parameters used in the Hamiltonian (3).

Figures 6(b) and (c) show $\sigma(\omega)$ for $\text{La}_8\text{Ca}_6\text{Cu}_{24}\text{O}_{41}$ and $\text{CuGeO}_3$ calculated using the $\text{Cu}_4\text{O}_{10}$ cluster. The structural data were taken from Siegrist et al.[37] at room temperature and Braden et al.[38] at $T = 295$ K, respectively. The effect of the buckling of the oxygen observed in the chain for $\text{La}_8\text{Ca}_6\text{Cu}_{24}\text{O}_{41}$ is taken into account as the average values of the deviation. The values of parameters used in the calculations are listed in Table I. For $\text{La}_8\text{Ca}_6\text{Cu}_{24}\text{O}_{41}$, we set $\varepsilon_\infty$ to be a rounded value of 3.4 between 3.5 for ladder systems and 3.3 for chain systems.[14] The value of $\Delta$ is, thus, smaller than that for $\text{Li}_2\text{CuO}_2$, although $\Delta V_M$ is very similar. A smaller CT gap of 2.1 eV is obtained compared to $\text{Li}_2\text{CuO}_2$. The CT gap of 3.2 eV for $\text{CuGeO}_3$ is remarkably larger than those for other cuprates (1.5 eV~2.3 eV). The origin of the large gap is the large $\Delta V_M$. The calculated $\Delta V_M$ is 52.2 eV and is larger than those for other cuprates, 46 eV $\leq \Delta V_M \leq$ 50 eV. Such an unusual value comes from the crystal structure of $\text{CuGeO}_3$ where the $\text{CuO}_2$ chains are tilting alternately about 60 degrees, and as a result Ge$^{4+}$ ions are very close to in-chain oxygen. The large gap is consistent with recent optical absorption data.[15,16]

### B. Magnetic Interactions and Magnetic Susceptibility

In this section, we estimate exchange interactions between Cu spins and examine magnetic properties for the edge-sharing CuO chain. An effective Hamiltonian describing the magnetic properties is the $S = 1/2$ Heisenberg model with NN and NNN exchange interactions ($J_1$-$J_2$ model, see Eq. (4)). $J_1$ is very sensitive to the Cu-O-Cu angle $\theta$. With approaching $\theta$ to 90 degrees from a larger value of $\theta$, the sign of $J_1$ changes from positive to negative, i.e. from AFM to FM, according to the Goodenough-Kanamori-Anderson rule. Therefore, it is necessary to consider the mechanisms that yield FM interaction between NN Cu spins. In Sec. II, we have introduced two types of the mechanism. One is due to Hund's coupling $K_p$ on oxygen ions which makes FM spin configuration favorable when two holes with $S = 1/2$ sit on different $2p_\alpha$ orbitals. Another is due to the direct exchange interaction between $3d_{xy}$ and $2p$ orbitals. If one spin is on the $3d_{xy}$ orbital and the other is on the $2p$ orbital which cannot overlap with the $3d_{xy}$ orbital, the two spins favor the parallel spin configuration due to $K_{pd}$. While the FM interactions induced by $K_p$ are of forth order of d-p hopping, those by $K_{pd}$ are of second order. Therefore, $J_1$ is more sensitive to $K_{pd}$ than $K_p$. The value of $K_{pd}$ used in the present study is 0.05 eV. This is within the region of values of $K_{pd}$ used in the previous studies.[38,39] By using this value, we can reproduce the experimental magnetic susceptibility obseved in $\text{Li}_2\text{CuO}_2$ as will be shown below.

We calculate $J_1$ and $J_2$ by using the exact diagonalization method for small clusters described by the Hamiltonian, Eq. (3). The parameter values are the same as those used in the calculation of $\sigma(\omega)$. $J_1$ and $J_2$ can be defined by the energy difference between the lowest singlet and the lowest triplet state for two-hole system of $\text{Cu}_2\text{O}_6$ and $\text{Cu}_2\text{O}_8$ clusters, respectively. The former cluster has two CuO$_4$ units sharing their edges, while the latter one has three edge-sharing CuO$_4$ units but without central Cu.[40] For $\text{Li}_2\text{CuO}_2$, we obtain $J_1 = -100$ K and $J_2 = 62$ K.

By using the calculated values of $J_1$ and $J_2$, we evaluate $\chi(T)$ performing numerical diagonalization for a 14-site $J_1$-$J_2$ chain. The result is plotted in Fig. 7(a) (solid line) together with experimental data along b-axis (square). The calculated result is in accord with exper-imental one in the wide range of temperature. As mentioned before, since $J_1$ is very sensitive to $K_{pd}$, we regard $K_{pd}$ as an adjustable parameter to be determined to obtain good agreement with experiment. Here, we note that $J_2$ is almost independent of $K_{pd}$, but strongly dependent on the hopping parameters $T_{dp_\alpha}^{ml}$ and $T_{p_\alpha p_\alpha}^{ll'}$. Since the good agreement is obtained by using a reasonable value of $K_{pd}$ without any change of hopping parameters, i.e. without changing $J_2$, we are convinced that the hopping parameters used in the calculation are reasonable and suitable for the description of the electronic states. At low temperatures below 30 K, the theoretical data show a peak and decrease with decresing temperature. This is due to the effect of frustration induced by $J_2$. However, for the quantitative comparison between theory and experiment at low temperatures, the larger cluster may be necessary to remove the size effect in the numerical calculation. In addition, the FM interchain interaction along the a-axis will give some contribution to the susceptibility.

In Fig. 7(a), numerical results for a free $S=1/2$ spin model and FM model ($J_1 < 0$, $J_2 = 0$) are shown for comparison. The result for the free spin model is in accord with the experimental data only at high temperature region ($T > 200$ K). In the case of the FM model ($J_1 = -10$ K), the curvature at around 100 K is larger than that in the experiment. For a larger value of $J_1$ ($= -100$ K), the curve of the susceptibility shift to higher temperature region and agreement at low temperatures becomes worse. From these comparisons, it is clearly found that $J_1$-$J_2$ model is most suitable for the description of $\chi(T)$ for $Li_2CuO_2$.

The values of $J_1$ ($J_2$) for $La_6Ca_8Cu_{24}O_{41}$ and $CuGeO_3$ are evaluated to be $-215$ K (78 K) and 140 K (18 K), respectively. $J_1$ for $La_6Ca_8Cu_{24}O_{41}$ is FM and stronger than that for $Li_2CuO_2$ ($-100$ K), while for $CuGeO_3$ $J_1$ is AFM. This trend is consistent with the systematic change of the Cu-O-Cu angle $\theta$ being 91, 94 and 99 degrees for $La_6Ca_8Cu_{24}O_{41}$, $Li_2CuO_2$ and $CuGeO_3$, respectively. $J_1$, $J_2$ and $\alpha$ are summarized in Table I.

It is known that the $J_1$-$J_2$ model contains various ground states depending on $\alpha$:[26] (i) When $J_1$ is positive (AFM) and $0 < \alpha < 0.241$, the ground state is a spin-liquid state without gap, and for $\alpha > 0.241$, it is in finite-gap phase.[28,29] In the case of $\alpha = 0.5$, the exact ground state is known to be a dimerized state, where the dimer singlets are formed between NNN spins.[27] (ii) When $J_1$ is negative (FM), depending on whether $-0.25 < \alpha < 0$ or $\alpha < -0.25$, the ground state is in FM phase or in a frustrated phase with zero magnetization. At the transition point ($\alpha = -0.25$), the exact ground state is known to be expressed as the state of products of all singlet bonds distributing uniformly on all lattice sites, so-called, uniformly-distributed-resonating-valence-bond state.[30] Our calculated $J_1$ and $J_2$ suggest that the ground state for $CuGeO_3$ is in gapless phase, while for $La_6Ca_8Cu_{24}O_{41}$ and $Li_2CuO_2$ are in the frustrated phase with zero magnetization.

The strength and the sign of $J_1$ are mainly determined by $\theta$, while the strength of $J_2$ is mainly determined by the distance between neighboring oxygens along the chain, i.e. the value of $T_{p_x p_x}^{ll'}$. $J_2$ for $La_6Ca_8Cu_{24}O_{41}$ is larger than that for $Li_2CuO_2$. This is because the distance of neighboring oxygens for $La_6Ca_8Cu_{24}O_{41}$, 2.76Å, is shorter than that for $Li_2CuO_2$, 2.86Å. For $CuGeO_3$, $J_2$ is much smaller than those for $La_6Ca_8Cu_{24}O_{41}$ and $Li_2CuO_2$. This small $J_2$ of $CuGeO_3$ is derived from two effects, i.e., (i) the longer distance between neighboring oxygens and (ii) the larger CT enegry as mentioned in Sec. III.

We have also calculated $\chi(T)$ for $La_6Ca_8Cu_{24}O_{41}$ and $CuGeO_3$, and compared them with experiment.[41,7] The results are shown in Fig. 7(b) and (c). The calculated results for $La_6Ca_8Cu_{24}O_{41}$ show excellent agreement with experimental data which are obtained subtracting contribution from the ladder layer.

On the other hand, the calculated results do not explain the experimental ones for $CuGeO_3$. It was suggested[18,19] that if the values $J_1$=160 K and $J_2$=58 K ($\alpha$=0.36) were used, the susceptibility could be explained. However, we have found that the value $J_2 \sim 60$ K causes $T_{p_x p_x}^{ll'}$ to be 1.5 times larger than that given above, keeping $\Delta$ the same. Then, such a large $T_{p_x p_x}^{ll'}$ results in the larger value of $J_2 \sim 200$K and $\sim 170$K for $La_6Ca_8Cu_{24}O_{41}$ and $Li_2CuO_2$, respectively. These values are not consistent with the experimental data of $\chi(T)$. For $CuGO_3$, it was also argued that the structual as well as phonon anomalies along the b-axis[46-49], the interchain interaction[42,44,43] and effects of Ge ions[45] bring about the anomalous magnetic properties[38]. These effects may be added into our $J_1$-$J_2$ model for the quantitative analyses of the magnetic properties.

## V. DISCUSSIONS AND CONCLUSIONS

In the previous section, we have examined the material dependence of the exchange interactions, $J_1$ and $J_2$. In order to summarize their dependences on $\theta$, we calculate $J_1$ and $J_2$ for several choices of parameter sets of $d_{Cu-O}$ and $\Delta$. The results are shown in Fig. 8. The thick and thin solid lines denote the results for $d_{Cu-O}$=1.90Å and 1.97Å, respectively, taking $\Delta$=3.0 eV as a typical value for the 1D cuprates. $d_{Cu-O}$=1.90Å and 1.97Å correspond to the lower and upper bounds of $d_{Cu-O}$ for the cuprates, respectively. The increase of $d_{Cu-O}$ causes the decrease of hopping amplitudes, resulting in the reduction of the magnitudes of $J_1$ and $J_2$. The dotted line denotes the result for $d_{Cu-O}$=1.95Å and $\Delta$=5.0 eV. The large $\Delta$ also reduces the magnitudes of $J_1$ and $J_2$ as expected. The bond angle minimizing $J_1$ deviates from 90 degrees. This is due to AFM contribution from direct O-O hoppings. We note that, if $T_{p_\alpha p_\alpha}^{ll'}$=0, the minimum appears just at $\theta = 90°$. In contrast to $J_1$, $J_2$ slightly decreases with increasing $\theta$. This stems from the increase of $d_{\text{O-O}}$ along the chains, i.e. the decrease of $T_{p_x p_x}^{ll'}$. For comparison, $J_1$ and $J_2$ for $\text{La}_6\text{Ca}_8\text{Cu}_{24}\text{O}_{41}$, $\text{Li}_2\text{CuO}_2$ and $\text{CuGeO}_3$ are plotted in the figure. For $\text{CuGeO}_3$, the large $\Delta$ is responsible for the deviation from the two solid lines. This figure may be useful for rough estimate of $J_1$ and $J_2$. For example, a compound $\text{Ca}_2\text{Y}_2\text{Cu}_5\text{O}_{10}$[50] ($\theta = 95°$, $d_{\text{Cu-O}} = 1.90\text{Å}$[51]) has $\Delta = 3.77$ eV, and is expected to show $J_1 \sim -20$ K and $J_2 \sim 50$ K. In fact, $J_1$ and $J_2$ are calculated to be $-25$ K and $55$ K from the cluster for $\text{Ca}_2\text{Y}_2\text{Cu}_5\text{O}_{10}$, respectively.

The 90° Cu-O-Cu bond structures are also seen in ladder materials as well as the edge-sharing chains. Those bond structures give interladder interactions. We can calculate these interactions, $J_{\text{int}}$, by the same procedure to get $J_1$. The calculated values of the interladder interactions for several materials are listed in Table II and are shown with the solid symbols in Fig. 8. Since the bond angle $\theta$ is nealy 90°, $J_{\text{int}}$ is FM and $\sim 300$ K, as espected from the lines in the figure. These interactions are not small as compared with intraladder interactions($\sim 1500$ K). Therefore, we should include them in the study of the magnetic properties.

In summary, we have studied systematically the electronic states and the magnetic interactions of the edge-sharing chains. In particular, we have focussed on $\text{La}_6\text{Ca}_8\text{Cu}_{24}\text{O}_{41}$, $\text{Li}_2\text{CuO}_2$ and $\text{CuGeO}_3$ in order to investigate the electronic states of this type of chain. The optical conductivity and the temperature depedence of the magnetic susceptibility have been measured for the sigle crystalline samples $\text{Li}_2\text{CuO}_2$ as a key material of the edge-sharing chain. In addition, in order to analyze the experimental data, the Cu-O cluster calculations have been performed by using the exact diagonalization method. It has been found that the spectral distribution of $\sigma(\omega)$ is in good accord with calculated one, showing the feature ascribed to the 90° Cu-O bond: The intensity of CT peak is strongly suppressed due to the orthogonality of the orbitals. The CT gap for $\text{CuGeO}_3$ has been estimated to be $\sim 3.2$ eV, which is much larger than those of other cuprates. This is consistent with recent optical measurements. This large gap is due to the large Madelung potential difference between Cu and O ions. As the important magnetic feature of the edge-sharing chain, there is the coexistence of (i) FM or AFM exchange interactions, $J_1$, between NN copper ions and (ii) AFM exchange interactions, $J_2$, between NNN copper ions. $J_1$ strongly depends on the bond angle $\theta$ between copper and oxygen ions, and changes from FM to AFM with increasing $\theta$. We have calculated $J_1$ and $J_2$ for the three cuprates from small Cu-O clusters in order to examine their material dependences. It has been found that the calculated $J_1$'s are FM for $\text{La}_6\text{Ca}_8\text{Cu}_{24}\text{O}_{41}$ and $\text{Li}_2\text{CuO}_2$, while AFM for $\text{CuGeO}_3$. By using $J_1$ and $J_2$ obtained, we could have reproduced the experimental data of $\chi(T)$ for $\text{Li}_2\text{CuO}_2$ and $\text{La}_6\text{Ca}_8\text{Cu}_{24}\text{O}_{41}$. However for $\text{CuGeO}_3$ the good argement has not been obtained. This disagreement indicates that the estimation of $J_1$ and $J_2$ from the fitting to the magnetic susceptibility within $J_1$-$J_2$ model for $\text{CuGeO}_3$ is not suitable. In order to remove the disagreement, it will be necessary to take into account the effects of Ge ion, interchain interaction and their temperature depencdences. Lasty, we have summarized the systematics of the exchange interactions generated by the 90° Cu-O-Cu bond. The dependences of $J_1$, $J_2$ and $J_{\text{int}}$ on the bond angle, the bond length and CT energy have been clarified.

## ACKNOWLEDGMENTS

We would like to thank H. Takagi, M. Kibune and M. Azuma for sharing with their experimental data prior to publication. This work was supported by a Grant-in-Aid for Scientific Research on Priority Areas from the Ministry of Education, Science, Sports and Culture of Japan. The parts of the numerical calculation were performed in the Supercomputer Center, Institute for Solid State Physics, University of Tokyo, and the supercomputing facilities in Institute for Materials Research, Tohoku University. Y. M. acknowledges the financial support of JSPS Research Fellowships for Young Scientists.

TABLE I. The parameters used in the cluster calculation and the calculated values of $J_1$ and $J_2$. Listed are the bond length between Cu and O ($d_{\text{Cu-O}}$ (Å)), the bond angle between Cu-O-Cu, the difference in Madelung site potential between Cu and O sites ($\Delta V_M$ (eV)), the charge-transfer energy ($\Delta$ (eV)) and the Slater-Koster hopping parameters for a hole (in units of eV). As for $(pp\pi)$, the relation of $(pp\pi) = -0.25(pp\sigma)$ is used. The calculated values of the NN exchange interaction ($J_1$ (K)), the NNN exchange interaction ($J_2$ (K)) and the ratio $\alpha(=J_2/J_1)$ are also listed together.

|             | $\text{La}_6\text{Ca}_8\text{Cu}_{24}\text{O}_{41}$ | $\text{Li}_2\text{CuO}_2$ | $\text{CuGeO}_3$ | $\text{Ca}_2\text{Y}_2\text{Cu}_5\text{O}_{10}$ |
|---|---|---|---|---|
| $d_{\text{Cu-O}}$ | 1.929 | 1.956 | 1.932 | 1.905 |
| $\theta$ | 91.4 | 94.0 | 99.1 | 95.0 |
| $\Delta V_M$ | 47.2 | 46.4 | 52.2 | 48.4 |
| $\Delta$ | 3.0 | 3.2 | 4.9 | 3.8 |
| $(pd\sigma)$ | 1.24 | 1.17 | 1.23 | 1.30 |
| $(pd\pi)$ | -0.57 | -0.54 | -0.56 | -0.60 |
| $(pp\sigma)$ | -0.62[a], -0.67[b] | -0.56[a], -0.69[b] | -0.52[a], -0.83[b] | -0.59[a], -0.77[b] |
| $J_1$ | -215 | -100 | 140 | -25 |
| $J_2$ | 78 | 62 | 18 | 55 |
| $\alpha$ | -0.36 | -0.62 | 0.13 | -2.2 |

[a]parallel to chain. [b]perpendicular to chain.

TABLE II. The parameters for the ladder materials. Listed are the bond length between Cu and O ($d_{\text{Cu-O}}$ (Å)), the bond angle between Cu-O-Cu, the difference in Madelung site potential between Cu and O sites ($\Delta V_M$ (eV)), the charge-transfer energy ($\Delta$ (eV)) and the calculated values of interchain and interladder interactions ($J_{\text{int}}$ (K)). For the ladder materials, $\epsilon_\infty$ is taken to be 3.5 in Eq. (3).

|             | $\text{SrCu}_2\text{O}_3$ | $\text{Sr}_{14}\text{Cu}_{24}\text{O}_{41}$ | $\text{La}_6\text{Ca}_8\text{Cu}_{24}\text{O}_{41}$ |
|---|---|---|---|
| $d_{\text{Cu-O}}$ | 1.969[a], 1.872[b] | 1.967[a], 1.882[b] | 1.972[a], 1.898[b] |
| $\theta$ | 87.4 | 88.6 | 89.5 |
| $\Delta V_M$ | 47.2[a], 47.1[b] | 47.4[a], 47.2[b] | 47.7[a], 47.0[b] |
| $\Delta$ | 2.6[a], 2.5[b] | 3.0[a], 3.0[b] | 3.1[a], 3.0[b] |
| $J_{\text{int}}$ | -358 | -303 | -262 |

[a]leg direction [b]rung direction

FIGURE CAPTIONS

FIG. 1. The crystall structures of (a) $La_6Ca_8Cu_{24}O_{41}$, (b) $CuGeO_3$ and (c) $Li_2CuO_2$. The edge-sharing chains are running along the directions of c-axis, c-axis and b-axis, respectively.

FIG. 2. The Cu-O structure of the edge-sharing chain. In (a), the black and white circles represent Cu and O ions, respectively, and $\theta$ is the bond angle between Cu-O-Cu. In (b) and (c), the orbital configurations for Cu and O ions are shown. The gray and white parts in the orbitals denote the phases with positive and negative, respectively. The axes of coordinates ($x' - y'$ axes) in (b) are taken along Cu-O bonds, while the axes of coordinates ($x - y$ axes) in (c) are rotated by 45 degrees from $x' - y'$ axes.

FIG. 3. Schematic description of the density of states for the CT insulator. LH: lower Hubbard band, UH: upper Hubbard band, NB: non-bonding $2p$ band, ZR: Zhang-Rice local singlet band. $E_F$ is Fermi energy. The labels **A** and **B** denote the processes of optical inter-band excitations.

FIG. 4. The experimental result of the optical conductivity spectrum $\sigma(\omega)$ with polarization parallel to the chain(b-axis) for single-crystalline sample of $Li_2CuO_2$.

FIG. 5. The experimental results of the magnetic susceptibility with magnetic field applied along the three principal axes. The inset represents the enlargement near the transition temperature.

FIG. 6. The calculated spectrum of the optical conductivity $\sigma(\omega)$ for (a) $Li_2CuO_2$, (b) $La_6Ca_8Cu_{24}O_{41}$ ( the contribution from chain ) and (c) $CuGeO_3$. The $\delta$ functions are convoluted with a Lorentzian broadening of 0.6 eV. $d_{Cu-O}$ and $e$ are bond length between Cu and O, and the elementary electric charge, respectively.

FIG. 7. The magnetic susceptibilities $\chi(T)$'s for a 14-site $J_1$-$J_2$ chain with $J_1$ and $J_2$ listed in Table I. (a): $Li_2CuO_2$, (b): $La_6Ca_8Cu_{24}O_{41}$ and (c): $CuGeO_3$. The squares in (a) and (b) denote the experimental data with magnetic field applied along chains. In the data for $La_6Ca_8Cu_{24}O_{41}$, we have substracted the contribution from ladder layers according to ref. 52. In (c), the experimental data for the three principal axes are shown, and the orbital susceptibility, estimated to be $10^{-4}$ emu/mole (taken from ref. 7), has been subtracted. In (a) and (b), the orbital susceptibility is assumed to be zero. In the calculations, $g$ factor has been taken to be 2.0.

FIG. 8. The dependences of $J_1$, $J_2$ and $J_{int}$ on $\theta$ for several choices of parameter sets of $d_{Cu-O}$ and $\Delta$. The squares, the circles and the triangles denote $J_1$ and $J_2$ for $La_6Ca_8Cu_{24}O_{41}$, $Li_2CuO_2$ and $CuGeO_3$, respectively. The crystal structures are taken from ref. 53 for $SrCu_2O_3$, ref. 54 for $Sr_{14}Cu_{24}O_{41}$ and ref. 37 for $La_6Ca_8Cu_{24}O_{41}$.

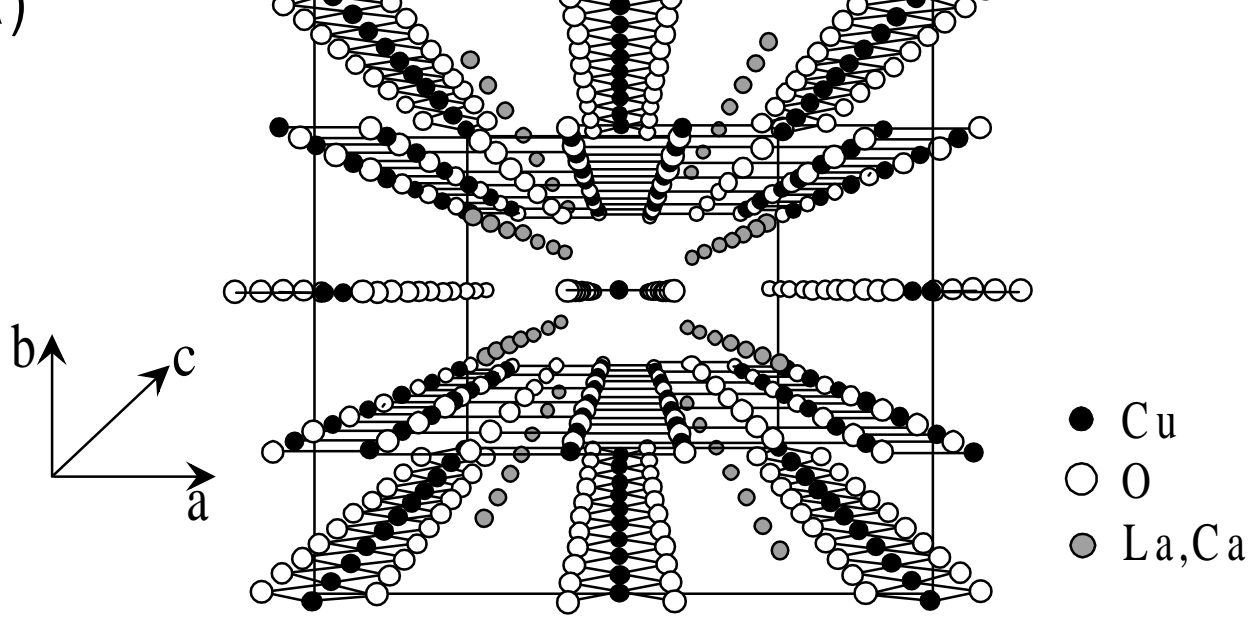

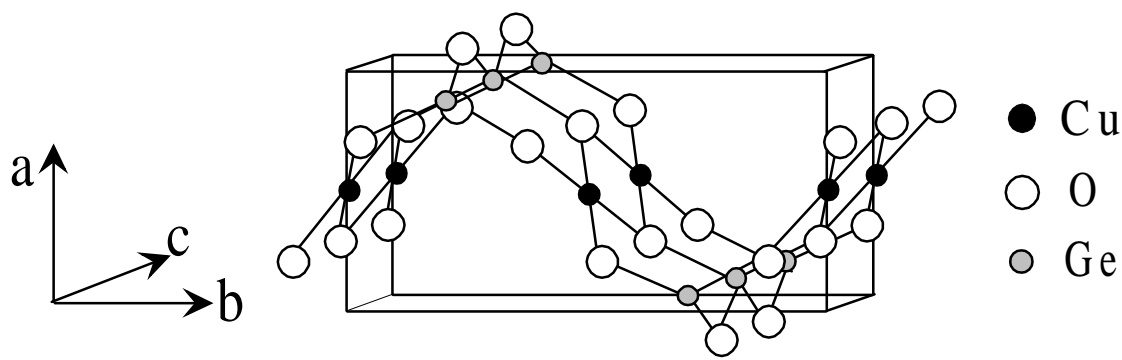

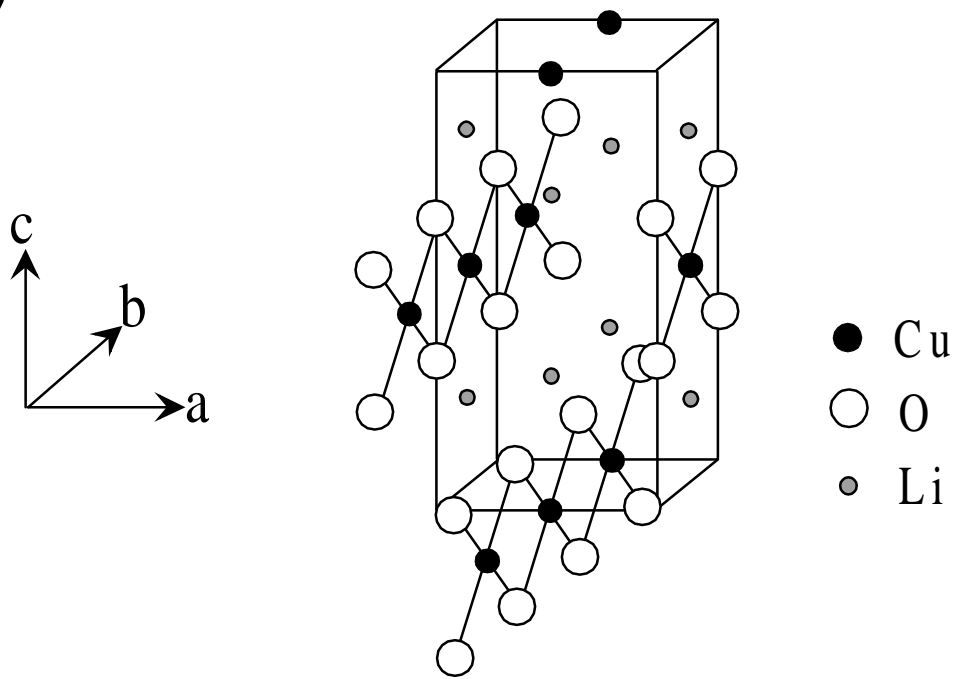

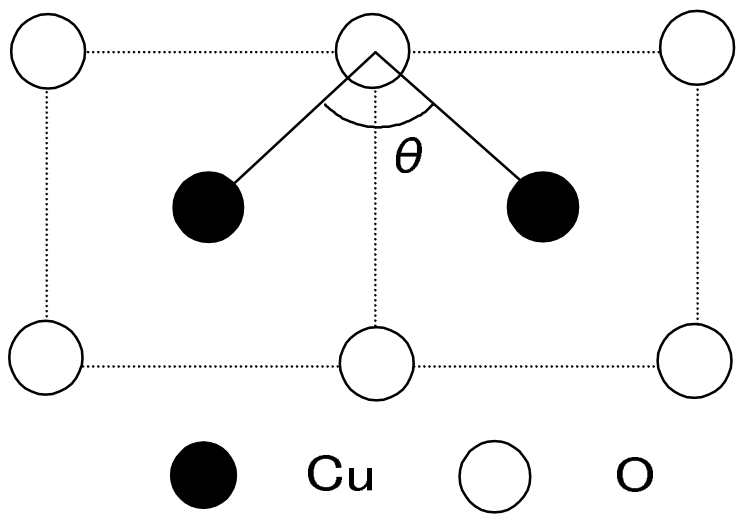

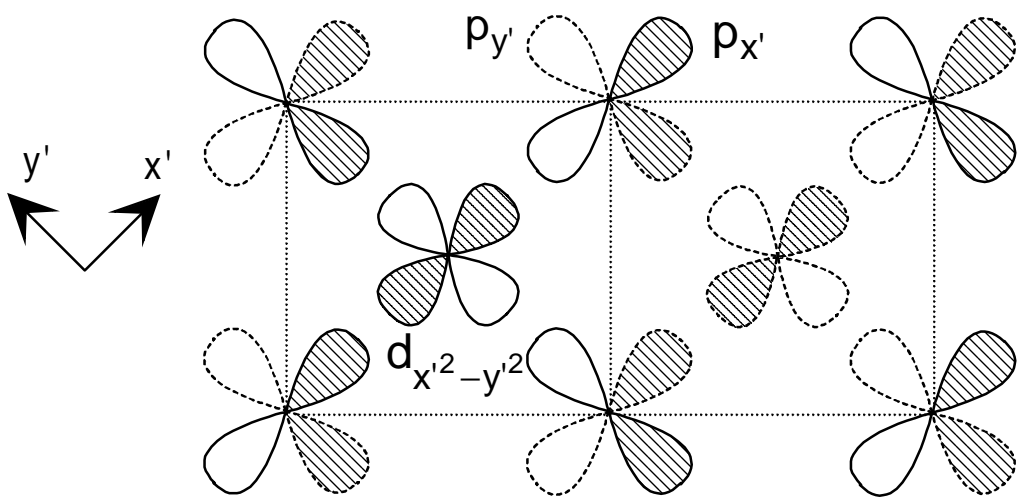

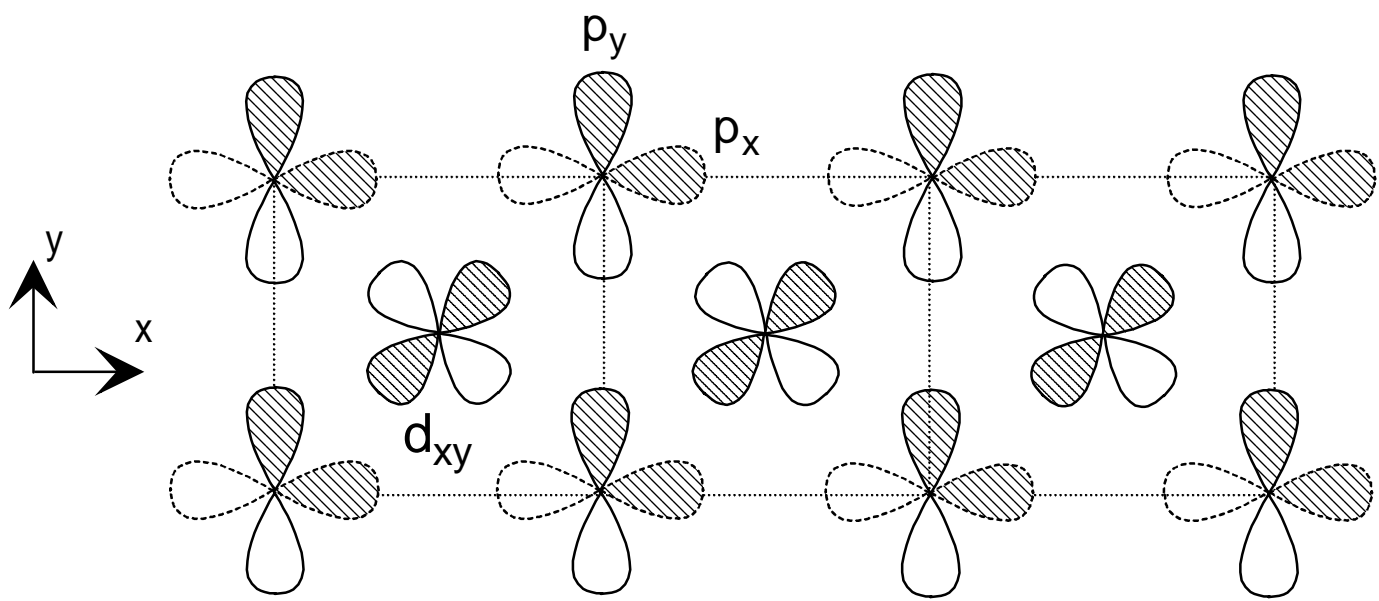

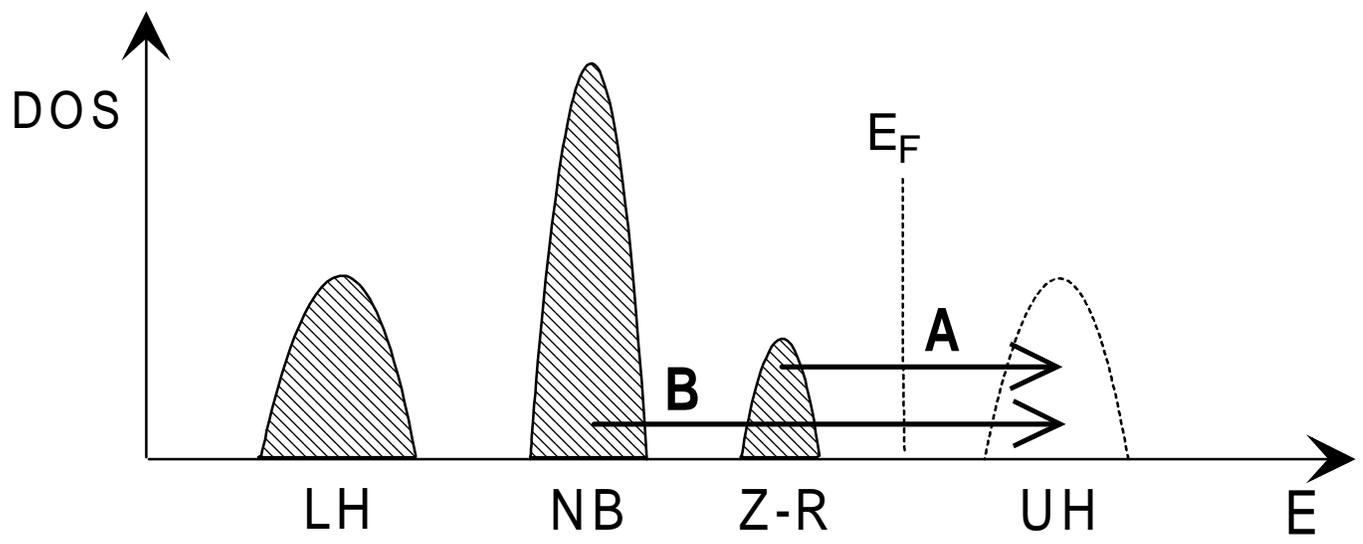

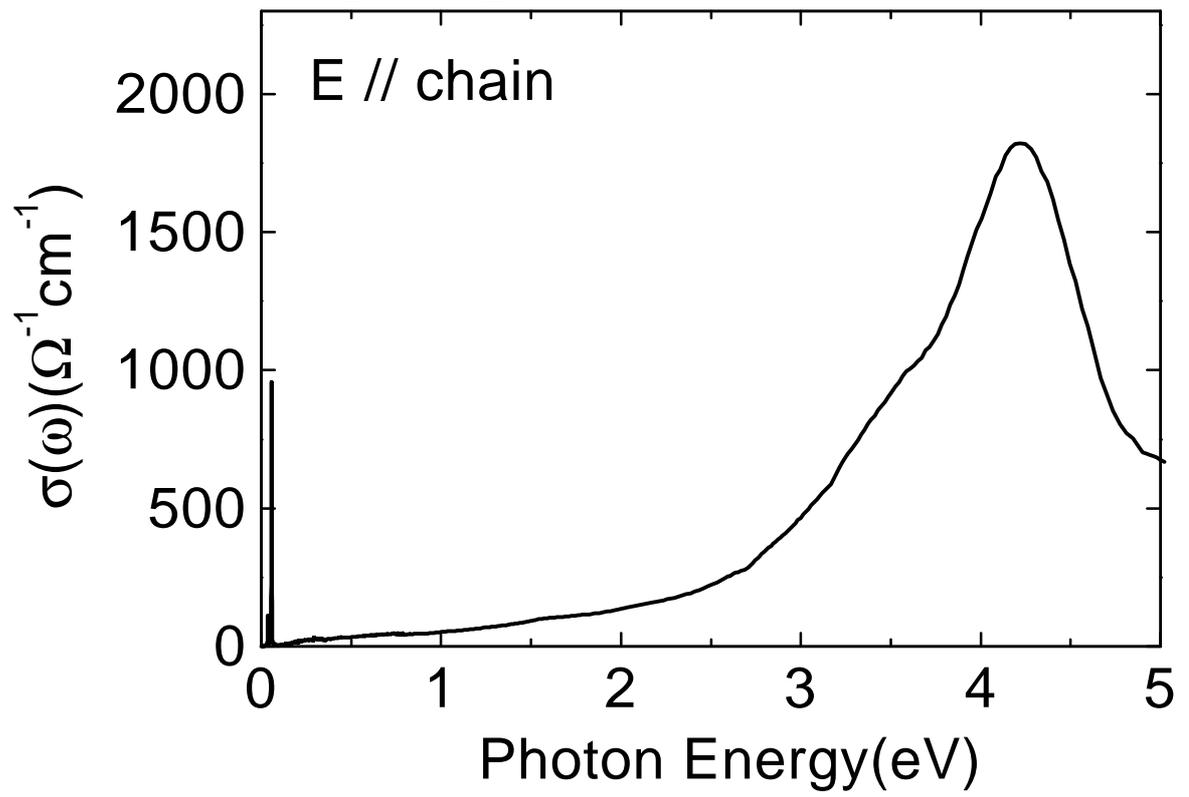

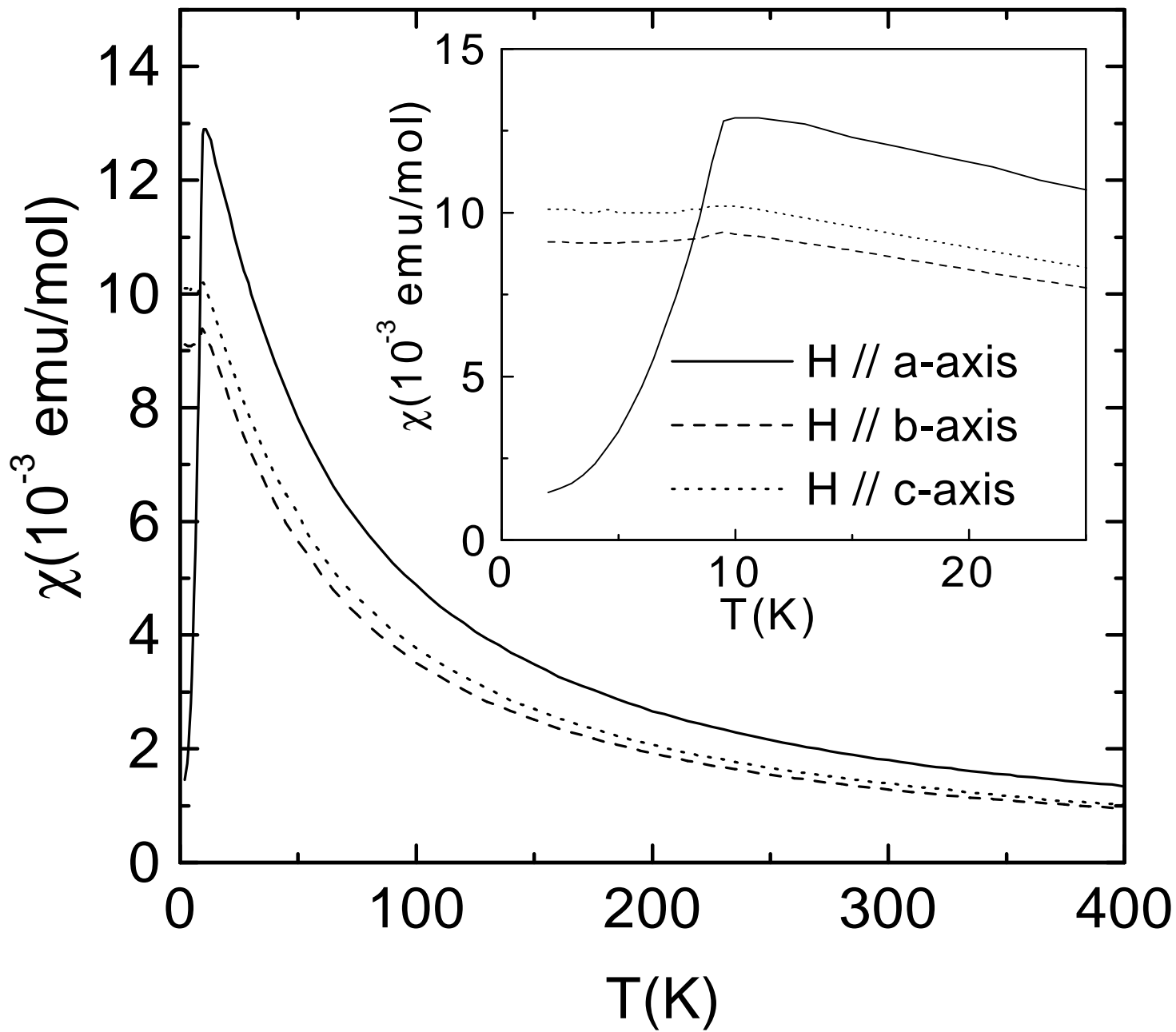

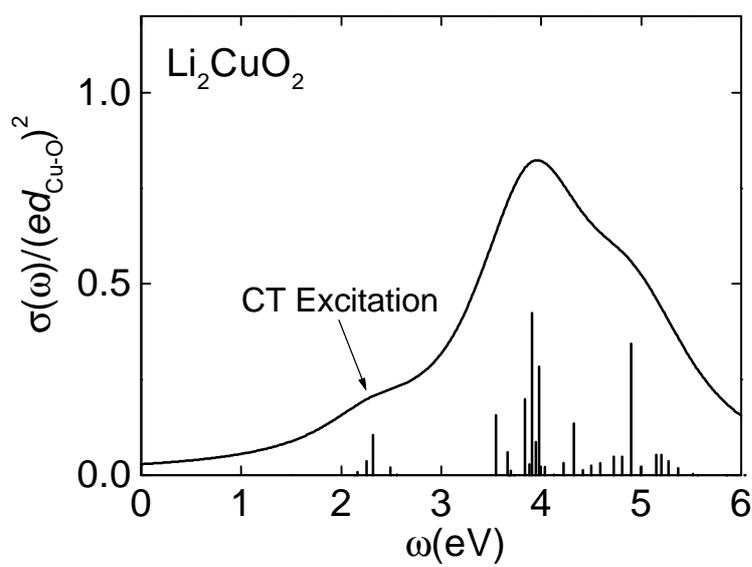

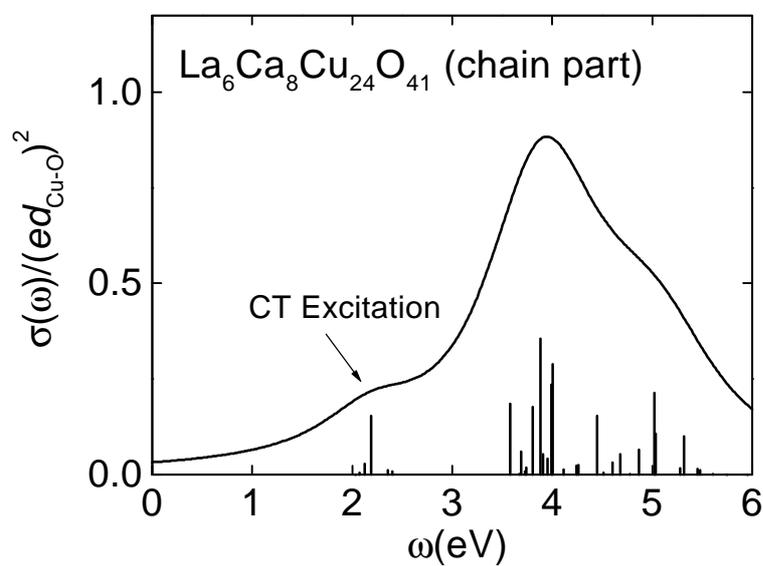

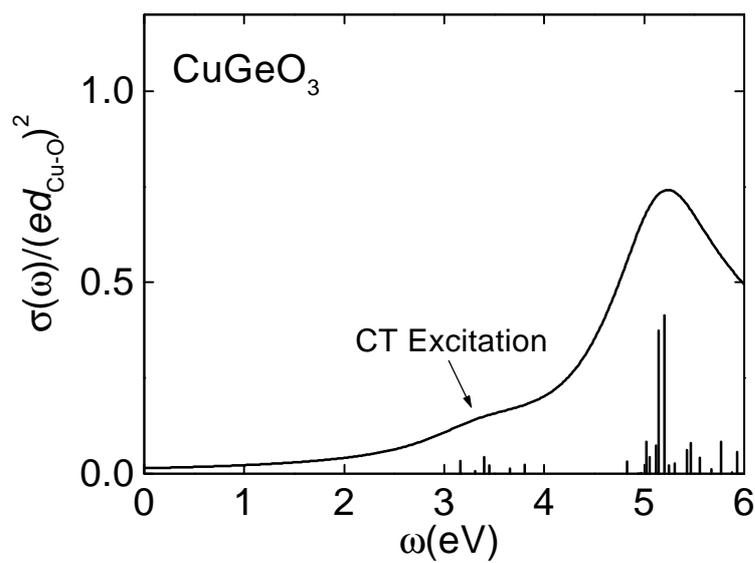

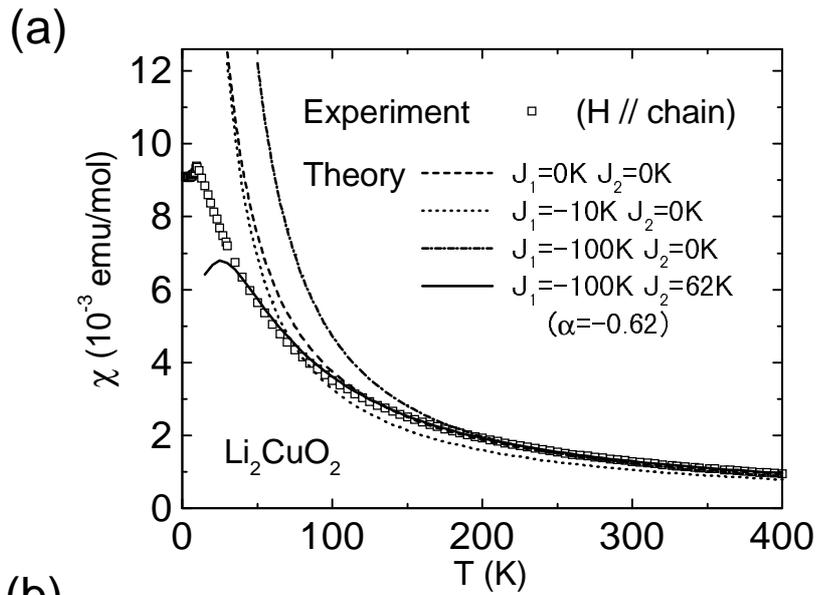

(a)

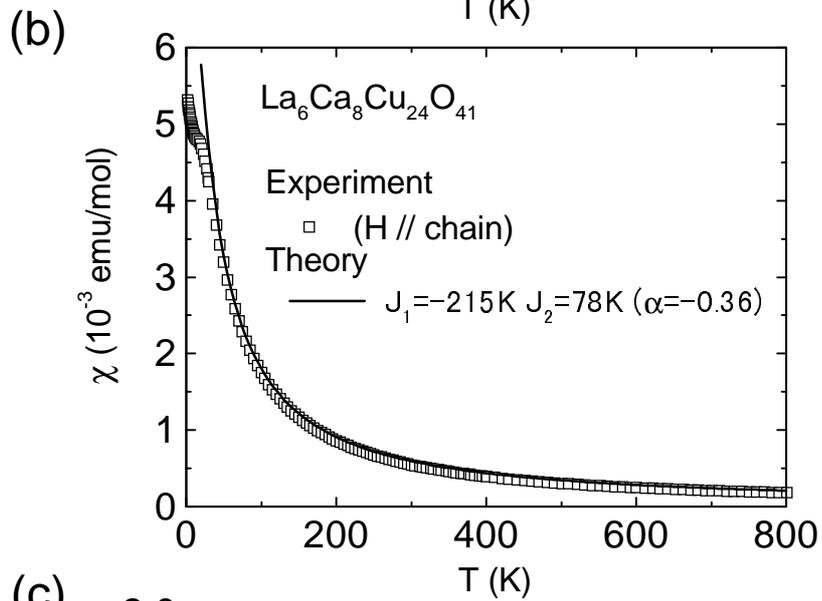

(b)

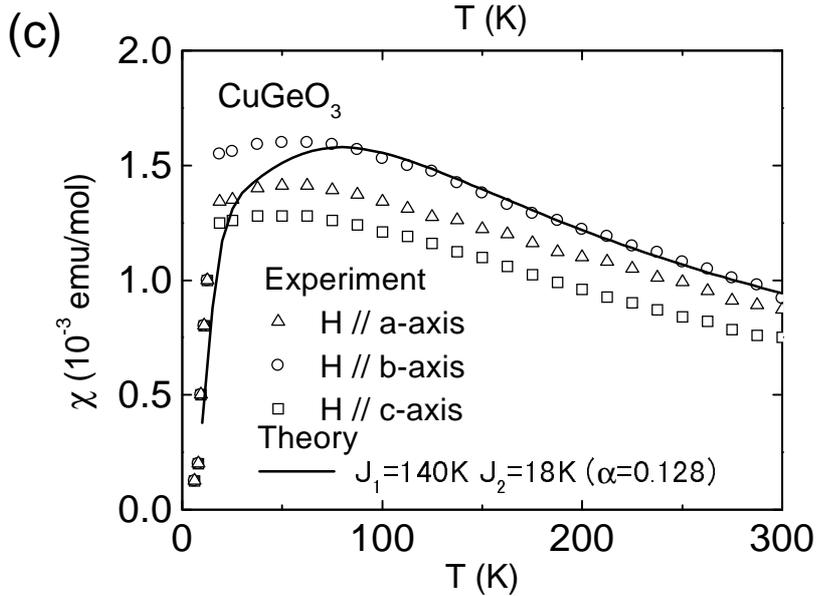

(c)

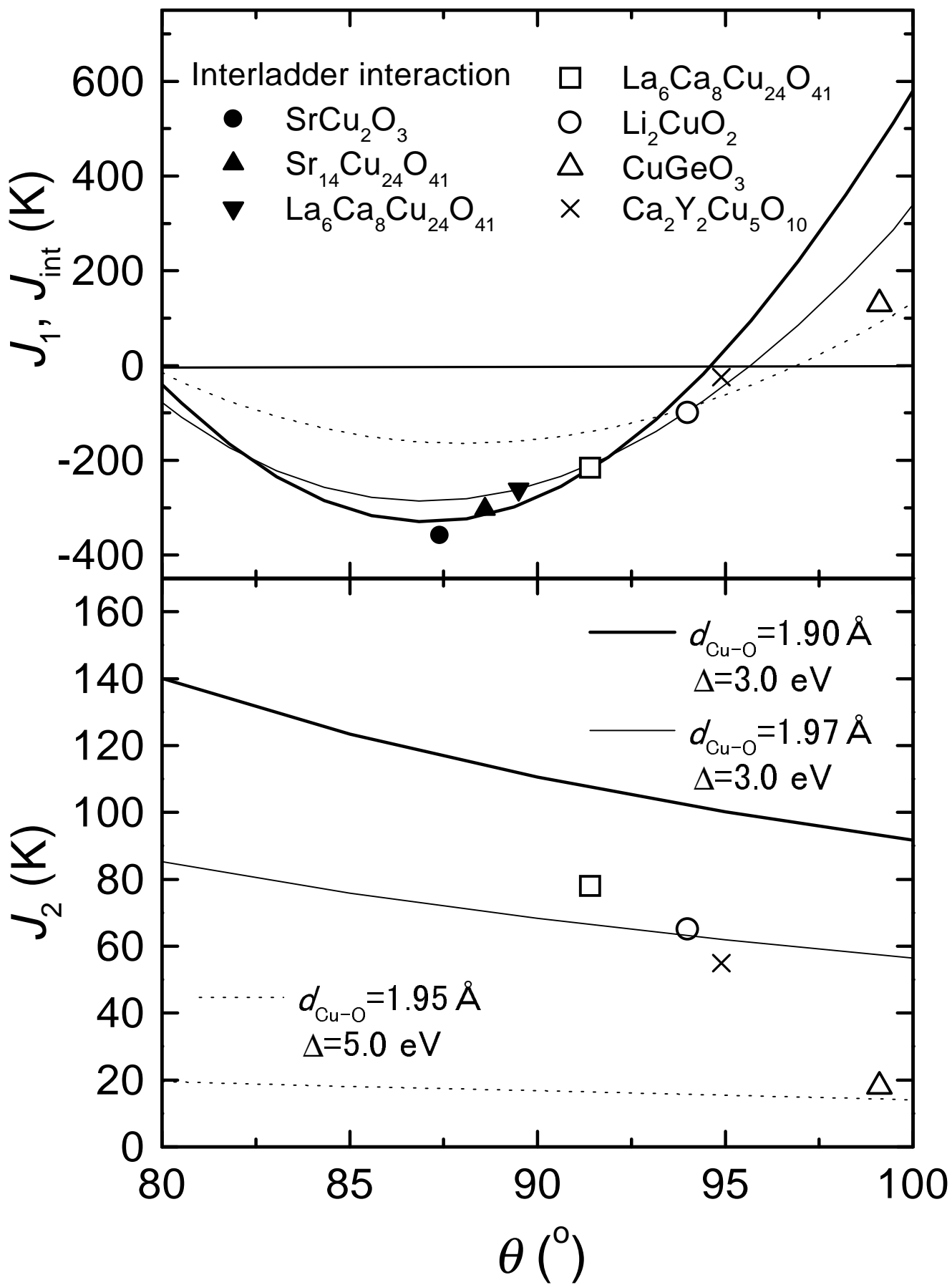